\documentclass[preprint,11pt,3p]{elsarticle}

\usepackage{hyperref}
\usepackage{graphicx}
\usepackage{graphics}
\usepackage{geometry}
\usepackage{subfig}
\usepackage{booktabs}
\usepackage{amsmath}
\usepackage{lscape}
\usepackage{subfig}
\usepackage{natbib}
\usepackage{lineno}
\modulolinenumbers[5]

\journal{Nuclear Instruments and Methods in Physics Research A}

\begin{document}
\begin{frontmatter}
\title{The TTIK approach for neutrons}

\author[texas]{V.\,Z.~Goldberg\corref{ksknote}}
\cortext[ksknote]{Corresponding author.}
\ead{vzgold@gmail.com }

\author[jinr,dsu]{E.\,M.~Gazeeva}
\author[jinr,dsu]{M.\,S.~Golovkov}
\author[jinr]{A.\,A.~Bezbakh}
\author[nla,spsu]{D.\,K.~Nauruzbayev}
\author[nla]{A.\,K.~Nurmukhanbetova}
\author[jinr,enu,inp]{Zh.~Kurmanaliyev}
\author[jinr,dsu]{A.~Serikov}
\author[jinr,warsaw]{B.~Zalewski}
\author[texas,texasdpa]{G.V.~Rogachev}

\address[texas]{Cyclotron Institute, Texas A\&M University, College station, Texas, 77843-3366 USA}
\address[jinr]{Joint Institute for Nuclear Research, Dubna, 141980 Russia}
\address[dsu]{Dubna State University, Dubna, 141982 Russia}
\address[nla]{Nazarbayev University, Nur-Sultan, 010000 Kazakhstan}
\address[spsu]{Saint Petersburg State University, Saint Petersburg, 198504 Russia}
\address[enu]{L.N. Gumilyov Eurasian National University, Nur-Sultan, 010000 Kazakhstan}
\address[inp]{Institute of Nuclear Physics, Almaty, 050032 Kazakhstan}
\address[warsaw]{Heavy Ion Laboratory, Warsaw, 02-093 Poland}
\address[texasdpa]{Department of Physics \& Astronomy Texas A\&M University, College station, Texas, 77843-3366 USA}

\begin{abstract}
We applied Thick Target Inverse Kinematics Method for a study of resonances decaying through neutron emission.
As a test we used a well-studied, because of its role in s-process in stars, ${}^{13}\mathrm{C}(\alpha,n)$ reaction.
The observed energy resolution for the ${}^{13}\mathrm{C}(\alpha,n)$ excitation function was $\sim$60~keV, the largest contributions coming from the time structure of the beam and the thickness of the neutron detector.
These measurements demonstrated the high efficiency of the approach and revealed a disagreement with R-matrix calculations based on parameters of the most recent previous analysis.
\end{abstract}

\begin{keyword}
Low energy resonance reactions, tick target inverse kinematic method, neutron detector
\PACS 21.10. -k \sep 25.40. Ny \sep 27.20. +n \sep 26.20. Kn \sep 29.30. -h
\end{keyword}

\end{frontmatter}


\section{Introduction}
\label{sec:intro}
About thirty years ago, one of us (V.Z.G.) proposed the Thick Target Inverse Kinematics (TTIK) method to search for $\alpha$-cluster states as scattering resonances in the interactions of heavy ions with  ${}^4\mathrm{He}$ \cite{ARTEMOV199052}.
In the TTIK technique, a beam of heavy ions is slowed in an extended helium gas target.
The light recoils, ($\alpha$-particles) are detected from a scattering event.
These recoils emerge from the beam ions-helium interaction and hit a Si detector array located at forward angles.
The beam ions are stopped in the gas, while the $\alpha$-particles, having smaller energy losses, are not.
Thus, measurement of ($\alpha$) recoil energies provides for a continuous excitation function of elastic scattering.
Due to inverse kinematics, zero degrees measurements correspond to 180${}^{\circ}$ $c.m.$ and it is at this angle that the best energy resolution can be achieved.
(In other studies, 25~keV $c.m.$ has been achieved \cite{NURMUKHANBETOVA2017125}).
Differences between the TTIK method and the classical approach are notable.
In the classical approach, using beams with well-defined energies and very thin targets, one can yield energy resolution less than 1~keV \cite{TOMPSON198042}.
However, the classical approach makes systematic fine step (commensurate the energy resolution) infeasible.
While the TTIK approach cannot compete with the classical approach in terms of energy resolution, it can provide the systematic excitation functions, which are impractical to obtain in the standard approach.
Initially, the TTIK approach has been proposed in experiments with stable beams, however, its greatest potential is for the investigations of resonance scattering induced by rare, exotic beams \cite{GOLDBERG2010307}.

The difference in the specific energy loss of a heavy ion beam and the recoils is a very important issue for the implementation of the TTIK approach.
The larger the difference, the better the resolution potentially achievable for the measurement of the excitation function.
As neutrons have no direct ionization loses, the application of the TTIK method to cases with neutrons in a final state is tantalizing.

The present work is the first application of TTIK method to measurements of excitation function of ${}^{13}\mathrm{C}(\alpha,n)$ reaction.
We used Time-of-Flight (ToF) technique (neutron detector time relative to the RF of the beam) for neutron energy measurements.
The choice of the reaction was based on the fact that the excitation functions for the ${}^{13}\mathrm{C}(\alpha,n)$ reaction have been studied at a broad range of $\alpha$-particle energies and because of the importance of this reaction for the nucleosynthetic s-process \cite{GALLINO1998}.
Additionally, our new results can be compared to a very comprehensive and recent analysis of the data relevant to this reaction, made by the Notre Dame University group \cite{HEIL2008}.
(This reference also contains a comprehensive review of previous studies.)

\section{Experiment}
\label{sec:Experiment}

\begin{figure}[htbp]
\centering
\includegraphics[width=0.8\linewidth]{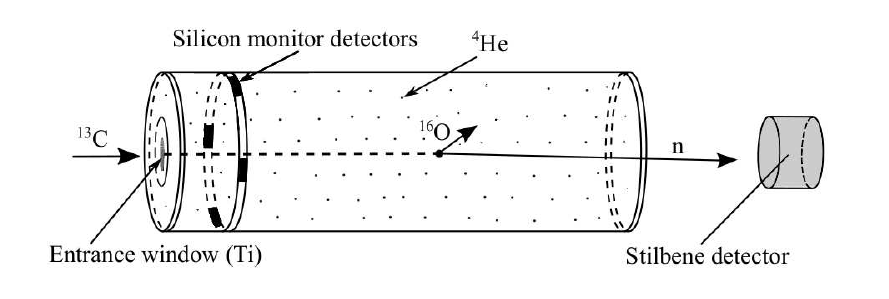}
\caption{Experimental layout.}
\label{fig:expSetup} 
\end{figure}

The experiment was performed at the DC-60 heavy-ion cyclotron in Nur-Sultan (Kazakhstan) using beam energies of 13.0 and 14.3~MeV.
The cyclotron parameters were tuned to provide a beam bunch with optimal time resolution at the expense of beam intensity.
The beam width was about 2 ns with $\sim$0.2\% energy resolution  at a beam intensity of a few nA.
The beam parameters were controlled by monitor detectors, described below.
The experimental setup is presented on Fig.~\ref{fig:expSetup}.
The cylindrically shaped scattering chamber (15.5 cm diameter and 53.0 cm length) was made of stainless steel with a wall thickness of 0.25 cm and filled with helium of 99.99\% purity.

The ${}^{13}\mathrm{C}$ beam from the cyclotron entered the scattering chamber through a thin entrance window made of 1.9~$\mu$m Ti foil.
Four monitor Si detectors were mounted on a ring-shaped holder inside the chamber in order and detected the ions elastically scattered from the entrance window.
The energy of the beam decreases as it proceeds through the gas.
While in most TTIK applications the unreacted beam is stopped in the gas, in this case, to increase the travel time of heavy ions between the resonances we used lower pressure than needed to range out the beam.
At the pressure of 25 torr, the beam ions deposit about 25\% of their energy in the gas and stop in a back  flange of the scattering chamber.
If the ${}^{13}\mathrm{C}(\alpha,n)$ reaction occurs, a neutron flying forward at near $0^\circ$ will pass through the solid angle subtended by a neutron detector  36~cm downstream, behind the back flange.

\begin{figure}[htbp] 
\centering
\includegraphics[width=0.7\linewidth]{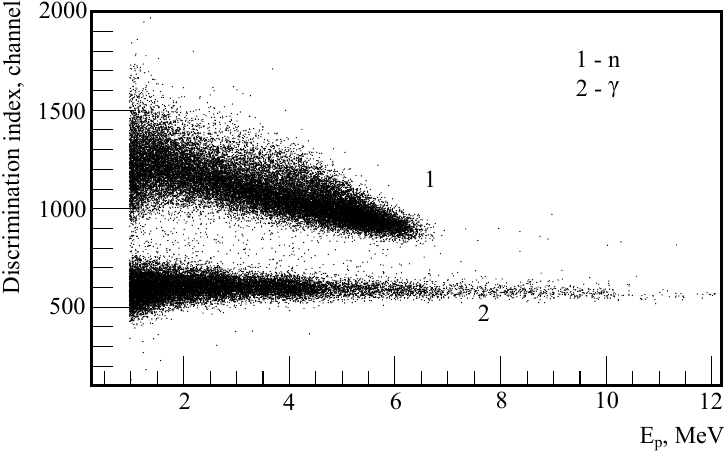}
\caption{n-$\gamma$ pulse shape discrimination.}
\label{fig:NeutrGamPulseShape}
\end{figure}

The neutron detector was a stilbene crystal of cylindrical shape (8 cm diameter, 5 cm thickness) optically coupled with Photonis XP 4312/B photomultiplier.
Stilbene has an excellent n-$\gamma$ separation ability which is exploited using standard pulse-shape discrimination (PSD) techniques.
The amplitude calibration was made with ${}^{60}\mathrm{Co}$ and ${}^{137}\mathrm{Cs}$ $\gamma$-sources.
Time resolution for amplitudes exceeding 0.5~MeV (equivalent electron energy) was better than 0.5~ns (FWHM).
Fig.~\ref{fig:NeutrGamPulseShape} demonstrates the quality of PSD made with MPD-4 module (Mesytec), where the horizontal axis shows the amplitude in recoil proton energy scale.
(Conversion from the electron to recoil proton scale  bases on the data of Ref.~\cite{ZHANG2013}.)
Following Ref.~\cite{ZHANG2013}, the amplitude of 0.5~MeV in electron scale corresponds to about $\sim$2.5~MeV in proton scale.

\begin{figure}[htb] 
\centering
\includegraphics[width=0.7\linewidth]{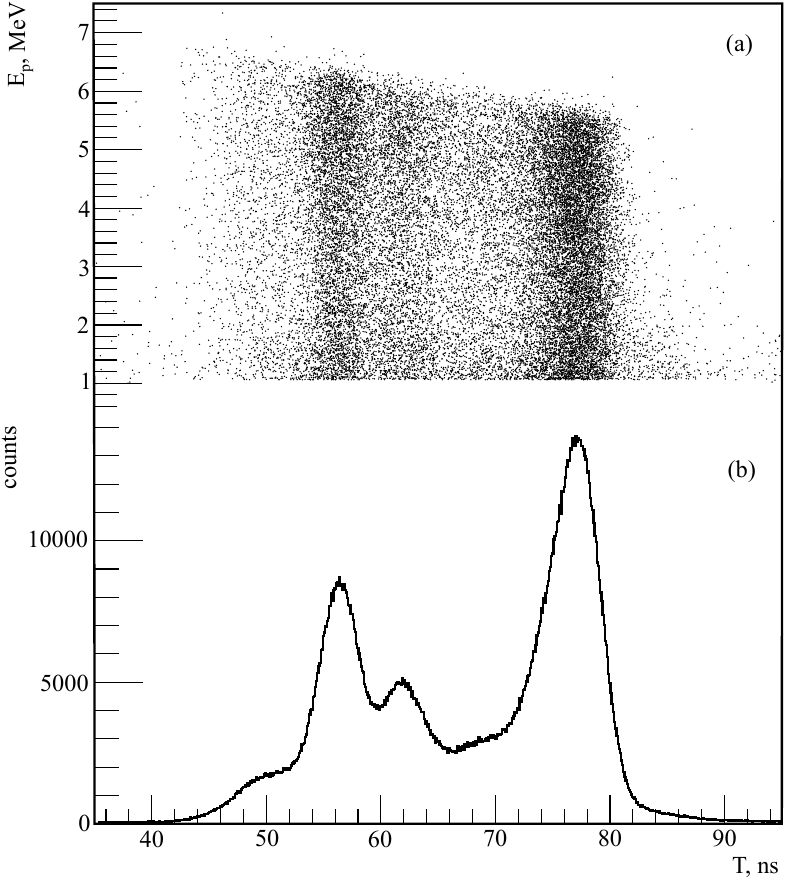}
\caption{(a) two dimensional $E_{p}$-T excitation function for $^{13}\mathrm{C}(\alpha,n)$ reaction and (b) projection of (a) spectrum on the X axis (see the text for explanations).}
\label{fig:NeutrAmplvsTime}
\end{figure}

A neutron detector spectrum is shown in Fig.~\ref{fig:NeutrAmplvsTime}.
The horizontal axis shows the neutron registration time relative to cyclotron RF and the vertical axis shows the amplitude in recoil proton energy scale.
The left (right) edge corresponds to the interaction of ${}^{13}\mathrm{C}$ near the entrance window (downstream end of the chamber).
The counts outside this interval cannot originate from the $\alpha-{}^{13}\mathrm{C}$ interaction and provide a measure of background.

In order to consider all the factors that influence the precision of the excitation function, a complete simulation of the experiment in GEANT4 toolkit was constructed.

\begin{figure}[htb] 
\centering
\includegraphics[width=0.9\linewidth]{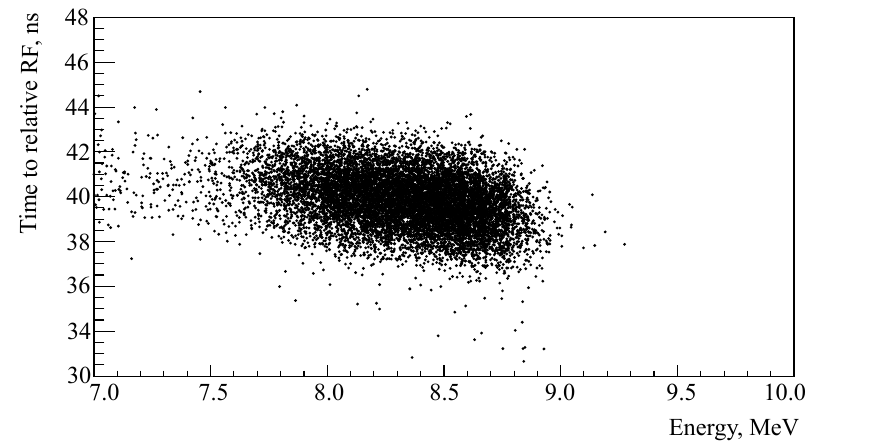}
\caption{Energy-time (E-T) spectrum of the ions scattered from the entrance window.}
\label{fig:ElasticEandT}
\end{figure}

\begin{figure}[htbp] 
\centering
\includegraphics[width=0.7\linewidth]{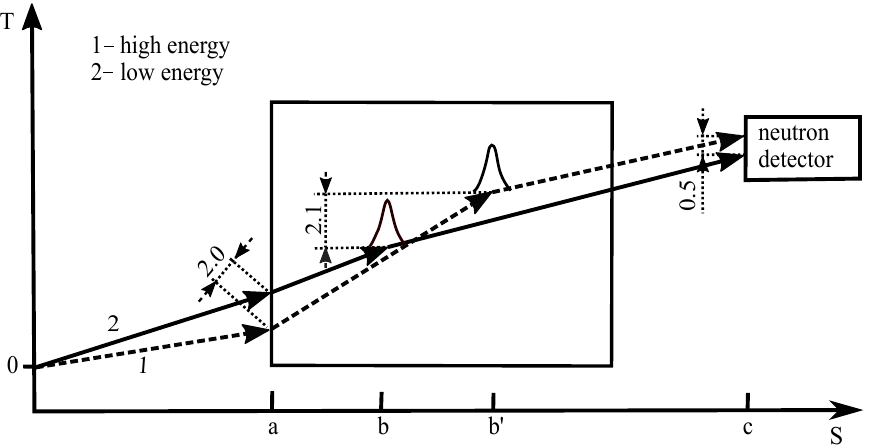}
\caption{
Diagram of the development of time spread of neutrons from (a) (see the text for explanations) 1(2) high (low) energy ions in the bunch, (a) 1 is ahead of 2 at the entrance to the scattering chamber, (b) resonance place for 2; (b') resonance place for 1, (c) 2 is ahead of 1 in the neutron detector (see the text for additional explanations)
}
\label{fig:LowHiEnergDiff}
\end{figure}

However before employing these simulations, it is instructive to obtain qualitative estimations of the time (energy) resolution using simple considerations.
Figures \ref{fig:ElasticEandT} and \ref{fig:LowHiEnergDiff} are provided to aid in this initial effort.
One of these elementary considerations is the time dispersion of the beam ions as they reach the chamber.
This dispersion is due to the initial energy spread of ions in the beam bunch and the flight path of 10~m between extraction from the cyclotron and the scattering chamber.
Fig.~\ref{fig:ElasticEandT} shows the energy-time (E-T) spectrum of the ions scattered from the entrance window observed in a monitor detector.
It is seen in Fig.~\ref{fig:ElasticEandT} that high-energy ions reach the detector earlier than the low-energy ions.
The time difference of approximately 2~ns is in agreement with a conventional cyclotron energy resolution ($\pm$~0.25\%) and the 10~m flight path.
The energy spread of the beam after passing through the window foil, calculated from the monitor detector spectrum, should be approximately 450~keV.
The main contribution to this spread is due to the thickness nonuniformity of the entrance window, estimated to be 9\%.
For reference, and to revert to time, beam ions lose approximately 0.45~MeV in 5.5~cm of gas and they need about 4.1~ns to travel this 5.5~cm distance.
The low-energy ions of the beam reach a resonance energy earlier, while the high-energy ions need to travel a longer distance (about 5.5~cm) and deposit an extra 0.45~MeV to reach the resonance energy.
Therefore, the neutrons from the decay of the narrow resonance in the ${}^{13}\mathrm{C}(\alpha,n)$ reaction will originate with a time spread of $\sim$~(4.1 -- 2.0)~ns, with the low-energy part of the beam reaching the resonance earlier.

Finally, the neutrons from the lowest energy part should travel 5.5~cm more than those from the high-energy part to the detector at zero degrees.
This difference amounts to about 1.6~ns.
(The neutron velocities are 2.5 higher than these of the heavy ions).
Summing up these effects: The faster ions from the machine get to the chamber sooner (about 2~ns), but they have to travel further (taking about 4.1~ns more) partially compensated by the reduced neutron flight time to the $0^\circ$ detector (about 1.6~ns).
The largely compensating effects lead to only an expected 0.5~ns time spread in the Time-of-Flight (ToF) of detected neutrons from the extreme ends of the initial energy distribution.
To evaluate the overall time resolution, we should also take into account the original width of beam bunch ($\sim$2~ns), and the neutron detector resolution of $\sim$1.6~ns (mainly from the finite thickness of the neutron detector).
Considering all three as independent factors yields 2.5~ns (FWHM).
This corresponds to a center-of-mass energy resolution of $\sim$60~keV.

\begin{figure}[htp] 
\centering
\includegraphics[width=0.7\linewidth]{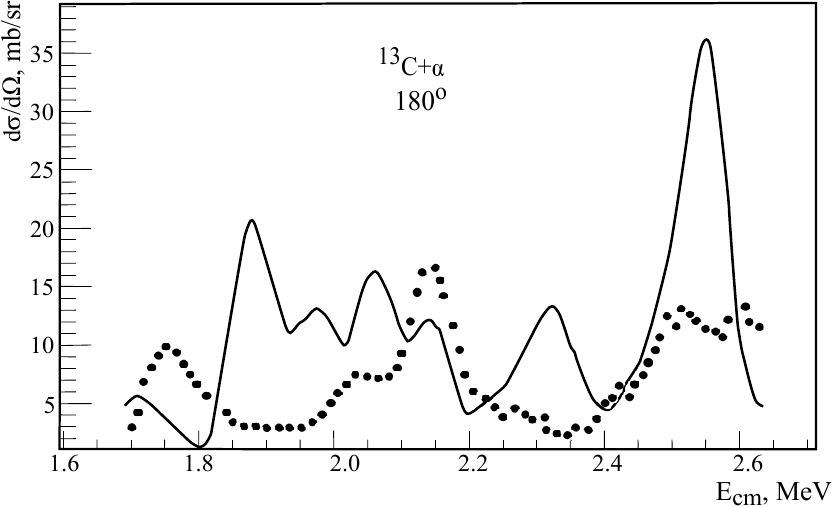}
\caption{The excitation function for the ${}^{13}\mathrm{C}(\alpha,n)$ reaction at 180${}^\circ$ c.m.
The bold curve is R-matrix calculations with parameters of Ref. \cite{HEIL2008}.}
\label{fig:ExtFunction180}
\end{figure}

\begin{figure}[!htp] 
\captionsetup[subfloat]{position=bottom}  
\centering
\subfloat[]{\includegraphics[width=0.7\linewidth]{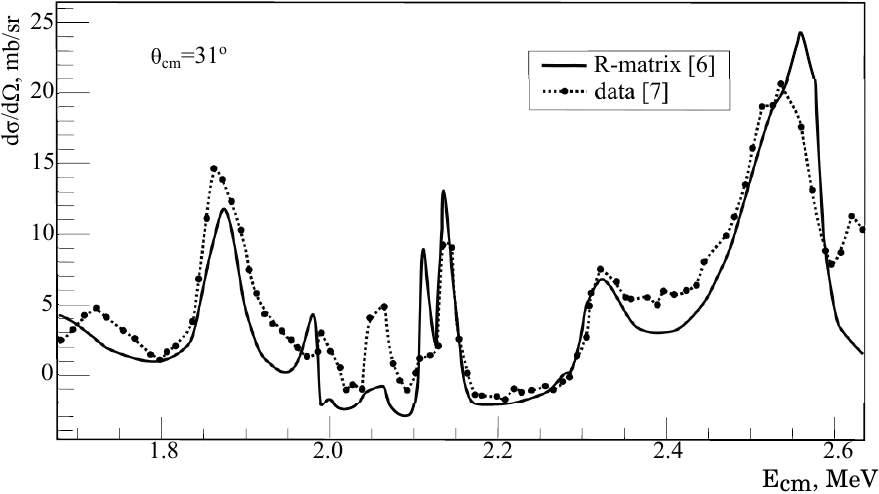}}
\vfill
\subfloat[]{\includegraphics[width=0.7\linewidth]{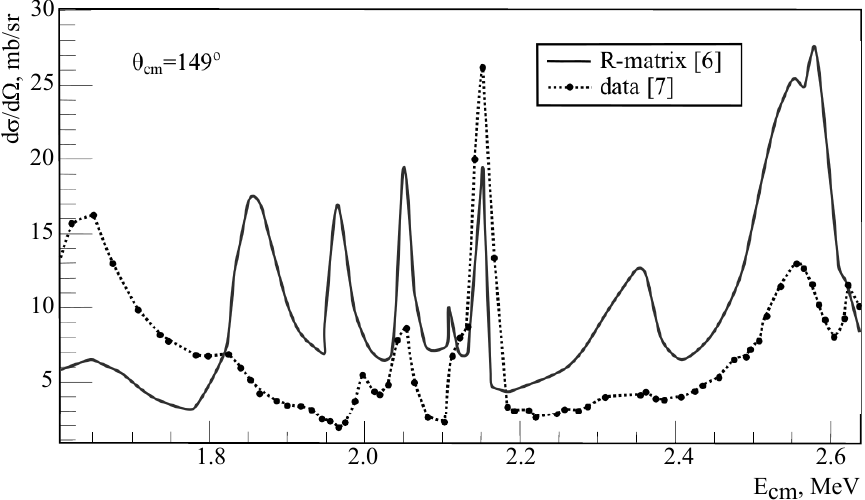}}
\caption{The excitation functions for the ${}^{13}\mathrm{C}(\alpha,n)$ reaction at 31${}^\circ$ and 149${}^\circ$ from Ref.~\cite{WALTON1957}.
The bold curves are R-matrix calculations with parameters of Ref.~\cite{HEIL2008}.}
\label{fig:ExtFunction149}
\end{figure}

Fig.~\ref{fig:ExtFunction180} demonstrates a combined spectrum using both the 13.0 and 14.3~MeV ${}^{13}\mathrm{C}$ measurements.
The energy resolution estimated by the FWHM resolution for the peak at 2.15~MeV $c.m.$ energy is $\sim$60~keV.
The solid line in Fig.~\ref{fig:ExtFunction180} shows R-matrix calculations with the parameters of Ref.~\cite{HEIL2008}.
The disagreement between these calculations and our measurements is evident.
Finding no obvious mistakes in either our work or that of the comprehensive work of \cite{HEIL2008}, we looked to older experimental data for additional insight.
The very old work by Walton et al. \cite{WALTON1957}, did in fact provide some insight.
These old excitation functions for the ${}^{13}\mathrm{C}(\alpha,n)$ reaction at symmetric forward and backward angles (31${}^\circ$ and 149${}^\circ$) together with R-matrix calculations with contemporary parameters~\cite{HEIL2008} are shown in Fig.~\ref{fig:ExtFunction149}.
Fair agreement is found at the forward angles but not at backward angles where the character of the disagreement is similar to that shown in Fig.~\ref{fig:ExtFunction180}.

\section{Discussion}
\label{sec:Discussion}

The present work shows that the energy resolution in the measurements of the excitation function for ${}^{13}\mathrm{C}(\alpha,n)$ reaction by the TTIK approach depends mainly upon the time width of the beam bunch (at the chamber) and on the time resolution of the neutron detector.
Both factors can be improved and we are planning on doing so.
We believe an energy resolution of $\sim$30 keV is achievable.
However, it is the high efficiency of this new approach that is most noteworthy.
The counting statistics in the observed peaks was well over $10^5$ counts.
These data resulted from only 30 hours measurement with a beam intensity of a few nA.

The present measurements, planned as a test, revealed an interesting result.
Evidently, the new result is related with the unusual geometry.
Obtaining equivalent data with the conventional approach would require neutron measurement at angles close to 180${}^\circ$, a difficult measurement.
The physical basis of the observed discrepancy likely resides in an unaccounted for interference of one or more broad (low spin) level(s) of opposite parity.
A manifestation of such levels, if they are broad, might be attributed to a background in an analysis of forward angle data.

If this is the case, the unaccounted levels might influence the calculations for the lower excitation energy region important for astrophysics.
However, a comprehensive analysis, like that in Ref.~\cite{HEIL2008} is needed to make a reliable conclusion.

It is not easy to predict the eventual applications of the specific TTIK approach described in the present work for the first time.
(This was also the case in the original TTIK work \cite{ARTEMOV199052}.)
One might imagine that the TTIK technique described here could be used to study  ($\alpha$,n) reactions on gas targets, many of which have not been thoroughly investigated.
This is largely due to experimental difficulties of working with gas targets in the conventional geometry.
The simplicity of the TTIK approach is attractive, even more so when the reactions are induced by low-intensity secondary beams.
In such cases, each beam particle would be tracked, removing a significant source of the ultimate center-of-mass energy uncertainty.
As a final point, measurements of resonant gamma rays delayed by a movement of low-energy heavy ions in an extended target seem interesting.
For such studies $\mathrm{LaBr_3}$ gamma, with their excellent (respectable) time (energy) resolution would be a good choice for these measurements.
As the first test gamma decays accompanying neutron (or proton) decays of the resonances to an excited state of a daughter nuclei might be selected because of the expected high yield.

\section{Conclusion}
\label{sec:Conclusion}

We performed the first measurements of the excitation function for the ${}^{13}\mathrm{C}(\alpha,n)$ reaction at 180${}^\circ$ $c.m.$ using TTIK method.
The excitation function was obtained in the $c.m.$ energy interval 1.7-2.6~MeV with energy resolution of $\sim$60~keV which is mainly defined by the width of the cyclotron beam burst and the neutron detector resolution.
The obtained excitation function is in general agreement with an old measurement in the backward hemisphere \cite{WALTON1957}.
However, the present results are in evident disagreement with a comprehensive fit \cite{HEIL2008} used to predict the ${}^{13}\mathrm{C}(\alpha,n)$  reaction rate needed for astrophysics.
We believe that the simplicity and high efficiency of the present approach will result in various applications.
In particular, it can be used for detection of $\gamma$-rays distributed in time according to resonant interaction of low energy heavy ions moving in extended target.

\section{Acknowledgement}
\label{sec:Acknowledgement}

The authors are indebted to Dr. S. Yennello for attention to the work and permanent support and to Dr. Lee Sobotka for reading the draft, for valuable comments and the discussion.
This work was supported in part by the Russian Science Foundation [grant number 17-12-01367].

Authors also acknowledge financial support from the Nazarbayev University [small grant
number 090118FD5346] and the Ministry of Education and Science of the Republic of
Kazakhstan [state-targeted program number BR05236454].

This work was supported by the U.S. Department of Energy, Office of Science, Office of Nuclear Science [award number DE-FG02-93ER40773] and by National Nuclear Security Administration through the Center for Excellence in Nuclear Training and University Based Research (CENTAUR)[grant number DE-NA0003841].
G.V.R. is also acknowledge the support of the Welch Foundation [grant number A-1853].

\bibliography{sections/biblio.bib}

\begin{thebibliography}{1}
\expandafter\ifx\csname url\endcsname\relax
  \def\url#1{\texttt{#1}}\fi
\expandafter\ifx\csname urlprefix\endcsname\relax\def\urlprefix{URL }\fi
\expandafter\ifx\csname href\endcsname\relax
  \def\href#1#2{#2} \def\path#1{#1}\fi

\bibitem{ARTEMOV199052}
K.~Artemov, O.~Belyanin, A.~Vetoshkin, Effective method of investigation of
  $\alpha$-cluster states, Soviet Journal of Nuclear Physics-Ussr 52 (1990)
  634--639.

\bibitem{NURMUKHANBETOVA2017125}
A.~Nurmukhanbetova, V.~Goldberg, D.~Nauruzbayev, G.~Rogachev, M.~Golovkov,
  N.~Mynbayev, S.~Artemov, A.~Karakhodjaev, K.~Kuterbekov, A.~Rakhymzhanov,
  Z.~Berdibek, I.~Ivanov, A.~Tikhonov, V.~Zherebchevsky, S.~Y. Torilov,
  R.~Tribble, Implementation of {TTIK} method and time of flight for resonance
  reaction studies at heavy ion accelerator {DC}-60, Nuclear Instruments and
  Methods in Physics Research Section A: Accelerators, Spectrometers, Detectors
  and Associated Equipment 847 (2017) 125--129.
\newblock \href {https://doi.org/10.1016/j.nima.2016.11.053}
  {\path{doi:10.1016/j.nima.2016.11.053}}.

\bibitem{TOMPSON198042}
W.~J. Thompson, J.~F. Wilkerson, T.~B. Clegg, J.~M. Feagin, E.~J. Ludwig,
  E.~Merzbacher, Atomic-excitation effects on nuclear reactions, Phys. Rev.
  Lett. 45 (1980) 703--706.
\newblock \href {https://doi.org/10.1103/PhysRevLett.45.703}
  {\path{doi:10.1103/PhysRevLett.45.703}}.

\bibitem{GOLDBERG2010307}
V.~Goldberg, B.~Roeder, G.~Rogachev, G.~Chubarian, E.~Johnson, C.~Fu,
  A.~Alharbi, M.~Avila, A.~Banu, M.~McCleskey, J.~Mitchell, E.~Simmons,
  G.~Tabacaru, L.~Trache, R.~Tribble, First observation of ${}^{14}${F},
  Physics Letters B 692~(5) (2010) 307--311.
\newblock \href {https://doi.org/10.1016/j.physletb.2010.07.054}
  {\path{doi:10.1016/j.physletb.2010.07.054}}.

\bibitem{GALLINO1998}
R.~{Gallino}, C.~{Arlandini}, M.~{Busso}, M.~{Lugaro}, C.~{Travaglio},
  O.~{Straniero}, A.~{Chieffi}, M.~{Limongi}, {Evolution and nucleosynthesis in
  low-mass asymptotic giant branch stars. {II}. neutron capture and the
  $s$-process}, The Astrophysical Journal 497 (1998) 388--403.
\newblock \href {https://doi.org/10.1086/305437} {\path{doi:10.1086/305437}}.

\bibitem{HEIL2008}
M.~Heil, R.~Detwiler, R.~E. Azuma, A.~Couture, J.~Daly, J.~G\"orres,
  F.~K\"appeler, R.~Reifarth, P.~Tischhauser, C.~Ugalde, M.~Wiescher, The ${
  }^{13}${C}$(\ensuremath{\alpha},n)$ reaction and its role as a neutron source
  for the $s$-process, Phys. Rev. C 78 (2008) 025803.
\newblock \href {https://doi.org/10.1103/PhysRevC.78.025803}
  {\path{doi:10.1103/PhysRevC.78.025803}}.

\bibitem{ZHANG2013}
X.~Zhang, X.~Yuan, X.~Xie, Z.~Chen, X.~Peng, J.~Chen, G.~Zhang, X.~Li, T.~Fan,
  G.~Zhong, L.~Hu, B.~Wan, A compact stilbene crystal neutron spectrometer for
  {EAST} {D-D} plasma neutron diagnostics, Review of Scientific Instruments
  84~(3) (2013) 033506.
\newblock \href {http://arxiv.org/abs/https://doi.org/10.1063/1.4797623}
  {\path{arXiv:https://doi.org/10.1063/1.4797623}}, \href
  {https://doi.org/10.1063/1.4797623} {\path{doi:10.1063/1.4797623}}.

\bibitem{WALTON1957}
R.~B. Walton, J.~D. Clement, F.~Boreli, Interaction of neutrons with oxygen and
  a study of the ${{\mathrm{C}}}^{13}(\ensuremath{\alpha},
  n){{\mathrm{O}}}^{16}$ reaction, Phys. Rev. 107 (1957) 1065--1075.
\newblock \href {https://doi.org/10.1103/PhysRev.107.1065}
  {\path{doi:10.1103/PhysRev.107.1065}}.

\end{thebibliography}

\end{document}